\documentclass[preprint,showpacs,preprintnumbers,amsmath,amssymb]{revtex4}

\usepackage{graphicx}
\usepackage{dcolumn}
\usepackage{bm}
\usepackage{color}
\usepackage{epsfig,subfigure,amsmath}

\begin{document}

\preprint{APS/123-QED}

\title{Metamaterial Loadings for Waveguide Miniaturization}


\author{H. Odabasi}
 \altaffiliation{Corresponding author. Email: odabasi.1@osu.edu}
\author{F. L. Teixeira}
\affiliation{ElectroScience Laboratory and Department of Electrical and Computer Engineering
\\The Ohio State University, Columbus Ohio, USA.}%

\date{\today}


\begin{abstract}

We show that a rectangular metallic waveguide loaded with metamaterial elements consisting of electric-field coupled (ELC) resonators placed at the side walls can operate well below the cutoff frequency of the respective unloaded waveguide. The dispersion diagrams indicate that propagating modes in ELC-loaded waveguides are of forward-type for both TE and TM modes. We also study the dispersion diagram and transmission characteristics of rectangular metallic waveguides simultaneously loaded with ELCs and split ring resonators (SRRs). Such `doubly'-loaded waveguides can support both forward wave and backward waves, and provide independent control of the propagation characteristics for the respective modes.

\end{abstract}

\pacs{41.20.-q, 41.20.Jb}


\maketitle

The study of metallic waveguides loaded with metamaterials has attracted much interest \cite{Marques_mm_wg_02a,Marques_mm_wg_02b,Kondratev_mm_wg_03,Marques_mm_wg_03,Falcone_mm_wg_03,Belov_mm_wg_05,Esteban_mm_wg_05,Nefedov_mm_wg_06,Xu_mm_wg_08,Hrabar_mm_wg_05,Baena_mm_wg_05,Lubkowski_mm_wg_06,Hrabar_mm_wg_06,Wu_mm_wg_07,Lubkowski_mm_wg_07,Carbonell_mm_wg_07,Antipov_mm_wg_07,Lubkowski_mm_wg_08,Iglesias_mm_wg_08,Teruel_mm_wg_09,Odabasi_josa,Meng_mm_wg_11,Dong_mm_wg_09,Xu_mm_wg_11, Chen_mm_wg_12,Ueda_mm_wg_05,Eshrah_mm_wg_05a,Eshrah_mm_wg_05b,Ueda_mm_wg_07,Edwards_mm_wg_08,Edwards_mm_wg_09}.  Because a rectangular waveguide behaves as an electric plasma for TE modes below cutoff~\cite{Marques_mm_wg_02a}, such waveguide can exhibit left-handed media behavior when loaded with negative permeability materials \cite{Marques_mm_wg_02a,Marques_mm_wg_02b,Kondratev_mm_wg_03,Marques_mm_wg_03}. In particular, waveguides loaded with split ring resonators (SRRs) providing negative permeability were studied extensively~\cite{Marques_mm_wg_02a,Marques_mm_wg_02b,Baena_mm_wg_05,Hrabar_mm_wg_05,Lubkowski_mm_wg_06,Hrabar_mm_wg_06,Wu_mm_wg_07,Lubkowski_mm_wg_07, Carbonell_mm_wg_07, Antipov_mm_wg_07, Lubkowski_mm_wg_08, Iglesias_mm_wg_08, Teruel_mm_wg_09, Meng_mm_wg_11}. One important feature of SRR-loaded waveguides is that they support propagation of backward waves.  
It is noteworthy that the anisotropic response of the metamaterial loadings plays a crucial role in the wave behavior~\cite{Kondratev_mm_wg_03,Marques_mm_wg_03,Hrabar_mm_wg_05,Belov_mm_wg_05,Xu_mm_wg_08,Meng_mm_wg_11}. In fact, this is precisely the reason why a waveguide with an isotropic negative permeability material loading does not support guided modes~\cite{Belov_mm_wg_05}. In order to produce backward wave propagation, the transverse magnetic permeability needs to be negative, whereas a negative longitudinal permeability produces forward waves. This was recently exploited for controlling backward and forward waves in SRR-loaded waveguides by means of a rotation on the SRR elements \cite{Meng_mm_wg_11}. Besides yielding left-handed behavior, metamaterial loadings also enable waveguide miniaturization \cite{Hrabar_mm_wg_05,Belov_mm_wg_05,Hrabar_mm_wg_06,Lubkowski_mm_wg_06,Odabasi_josa,Meng_mm_wg_11}. 
Though most of the research has focused on SRR-loaded waveguides, other types of metamaterial loadings are also effective~\cite{Ueda_mm_wg_05,Eshrah_mm_wg_05a,Eshrah_mm_wg_05b,Belov_mm_wg_05,Esteban_mm_wg_05,Nefedov_mm_wg_06,Lubkowski_mm_wg_06,Lubkowski_mm_wg_07, Ueda_mm_wg_07, Xu_mm_wg_08, Lubkowski_mm_wg_08, Dong_mm_wg_09}. In \cite{Belov_mm_wg_05}, it was shown that a periodical array of resonant scatterers can provide a passband below cutoff. Dispersion properties of waveguides loaded with equivalent electric and magnetic dipoles were studied for different dipole orientations~\cite{Belov_mm_wg_05}. Loadings such as ferrites and dielectric-filled corrugations were also proposed for realization of left-handed media~\cite{Ueda_mm_wg_05,Eshrah_mm_wg_05a,Eshrah_mm_wg_05b,Ueda_mm_wg_07}. 

It is also of interest to study what happens when a metallic waveguide is loaded with negative permittivity media. It can be easily shown that negative permittivity does not produce a passband for TE modes. On the other hand, by invoking the duality between TE and TM modes, rectangular waveguides behave as magnetic plasma for TM modes below cutoff~\cite{Esteban_mm_wg_05,Nefedov_mm_wg_06,Xu_mm_wg_08}. Hence, when loaded with negative permittivity media, they mimic left-handed media. In particular, wire media was proposed to enable backward waves for TM modes~\cite{Esteban_mm_wg_05,Nefedov_mm_wg_06}. Experimental results on backward waves for TM modes were presented in~\cite{Xu_mm_wg_08}.  

In this work, we investigate TE and TM modes in waveguides loaded with electric-field-coupled (ELC) resonators. These resonators were proposed~\cite{Schurig_mm_elc_06} to yield negative permittivity, as an alternative to wire media. ELC has the advantage of easily tunable response and can be excited under different polarizations, enabling different resonant frequencies and characteristics. This feature of ELC was recently exploited for the design electrically-small antennas~\cite{Odabasi_mm_antenna_13}, in conjunction with complementary-electric-field-coupled (CELC) resonators~\cite{Hand_mm_celc_08}. 

Most of the prior work on metamaterial-loaded waveguides has considered a `homogeneous' type of loading, i.e., where one single element type is used. In this work, we also show that waveguides simultaneously loaded with ELC and SSR resonators can support both backward and forward modes below the original cutoff frequency. In this case, each resonator type can provide independent control of the respective propagating modes.
\begin{figure}[htbp]
\centering
\includegraphics[width=0.58\textwidth]{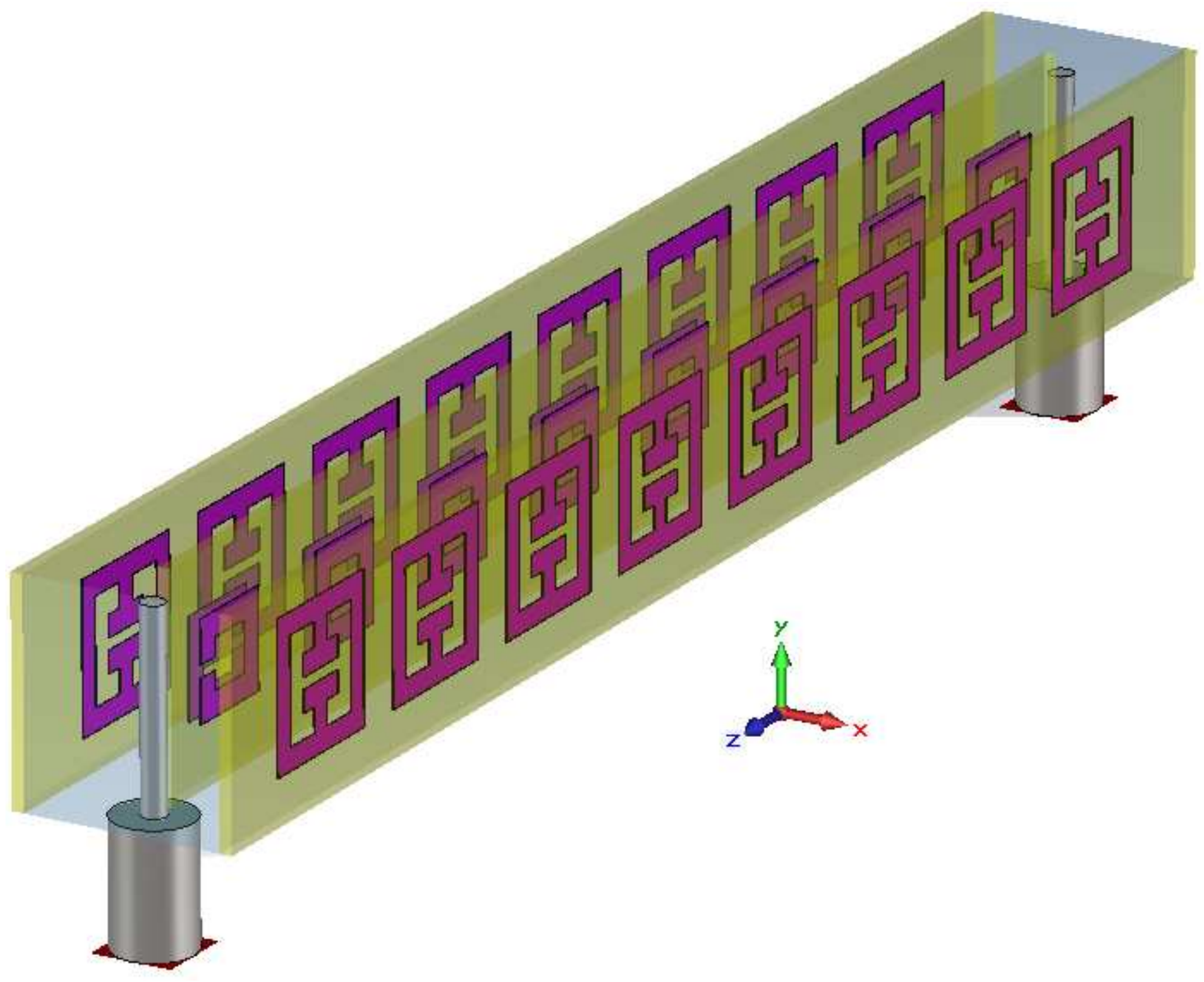}
\caption{Configuration of a (ELC+SRR)-loaded rectangular waveguided. The ELC resonators are placed at the lateral walls, while the SRR resonators are placed at the center of the waveguide.}
\label{fig1}
\end{figure}
\begin{figure}[htbp]
\centering
\subfigure[]{\includegraphics[width=0.42\textwidth]{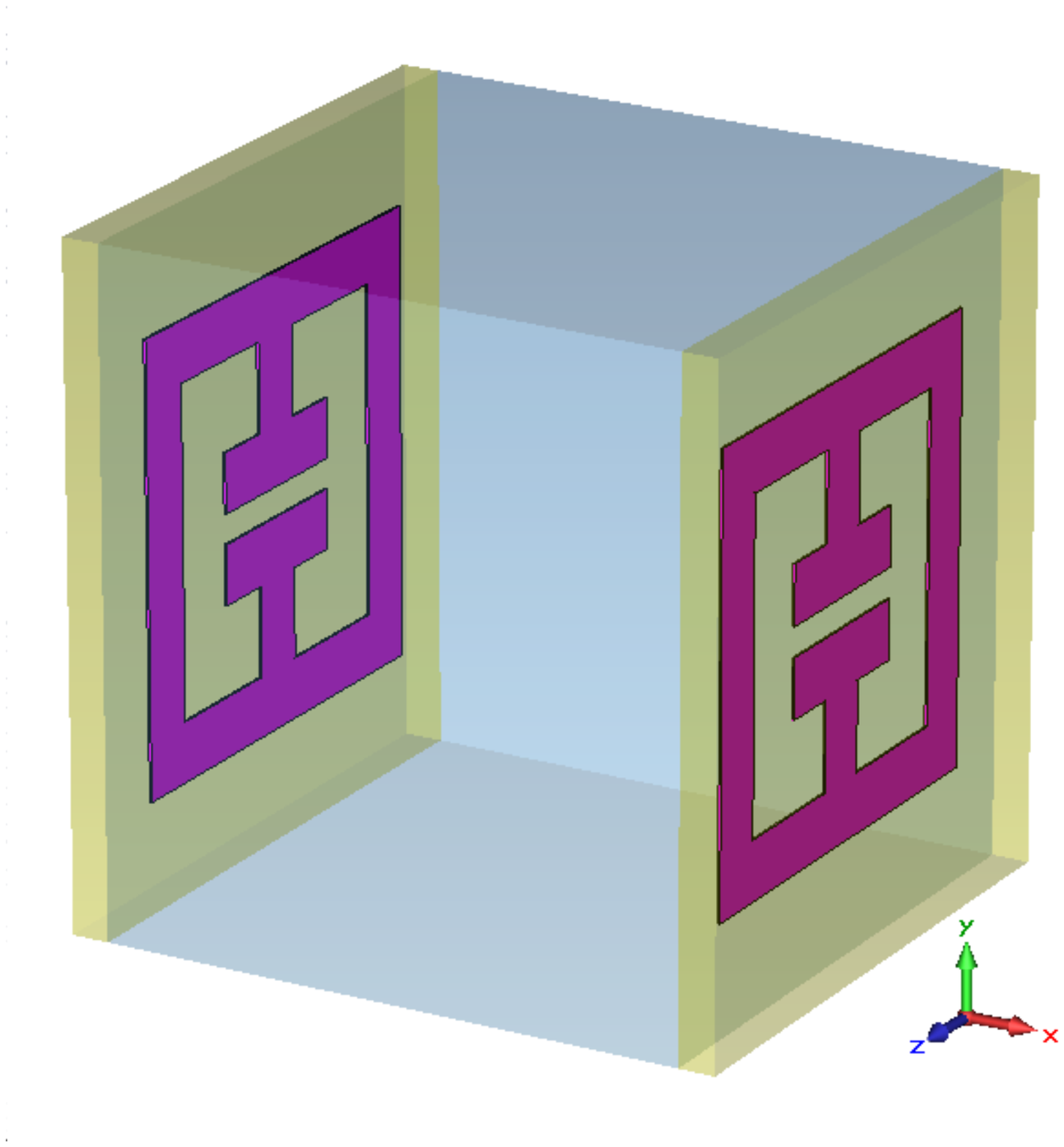}}
\subfigure[]{\includegraphics[width=0.42\textwidth]{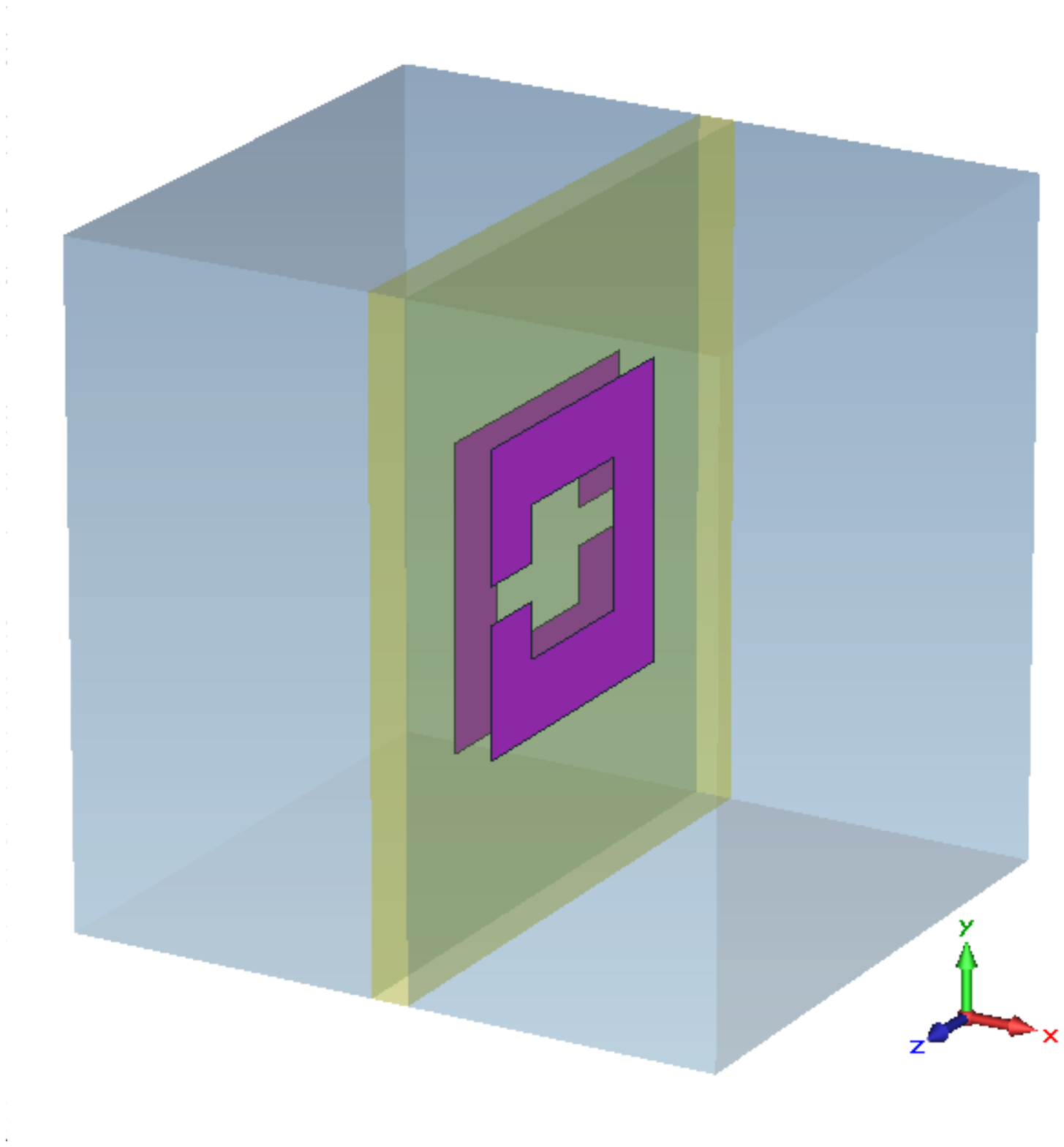}}
\subfigure[]{\includegraphics[width=0.42\textwidth]{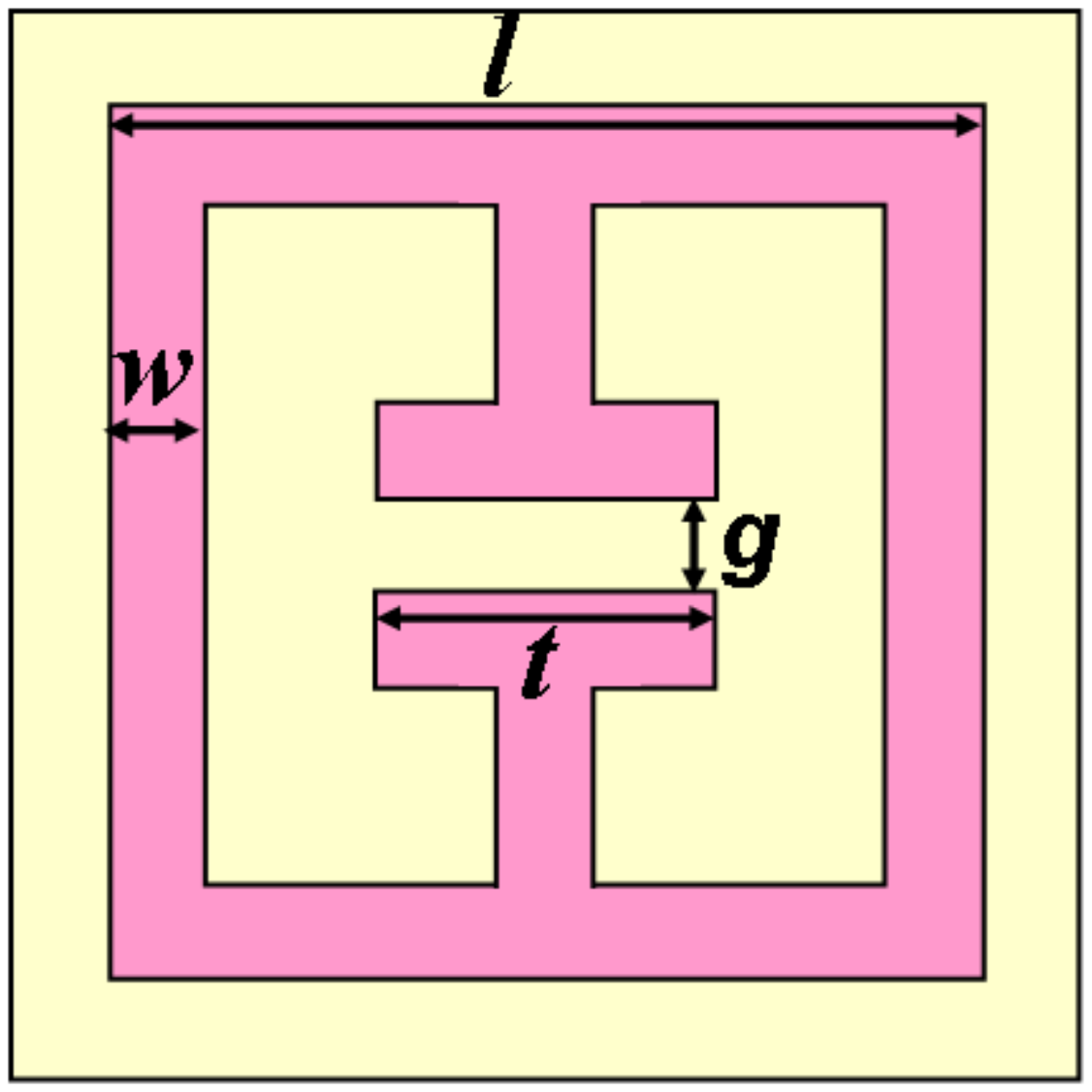}}
\subfigure[]{\includegraphics[width=0.42\textwidth]{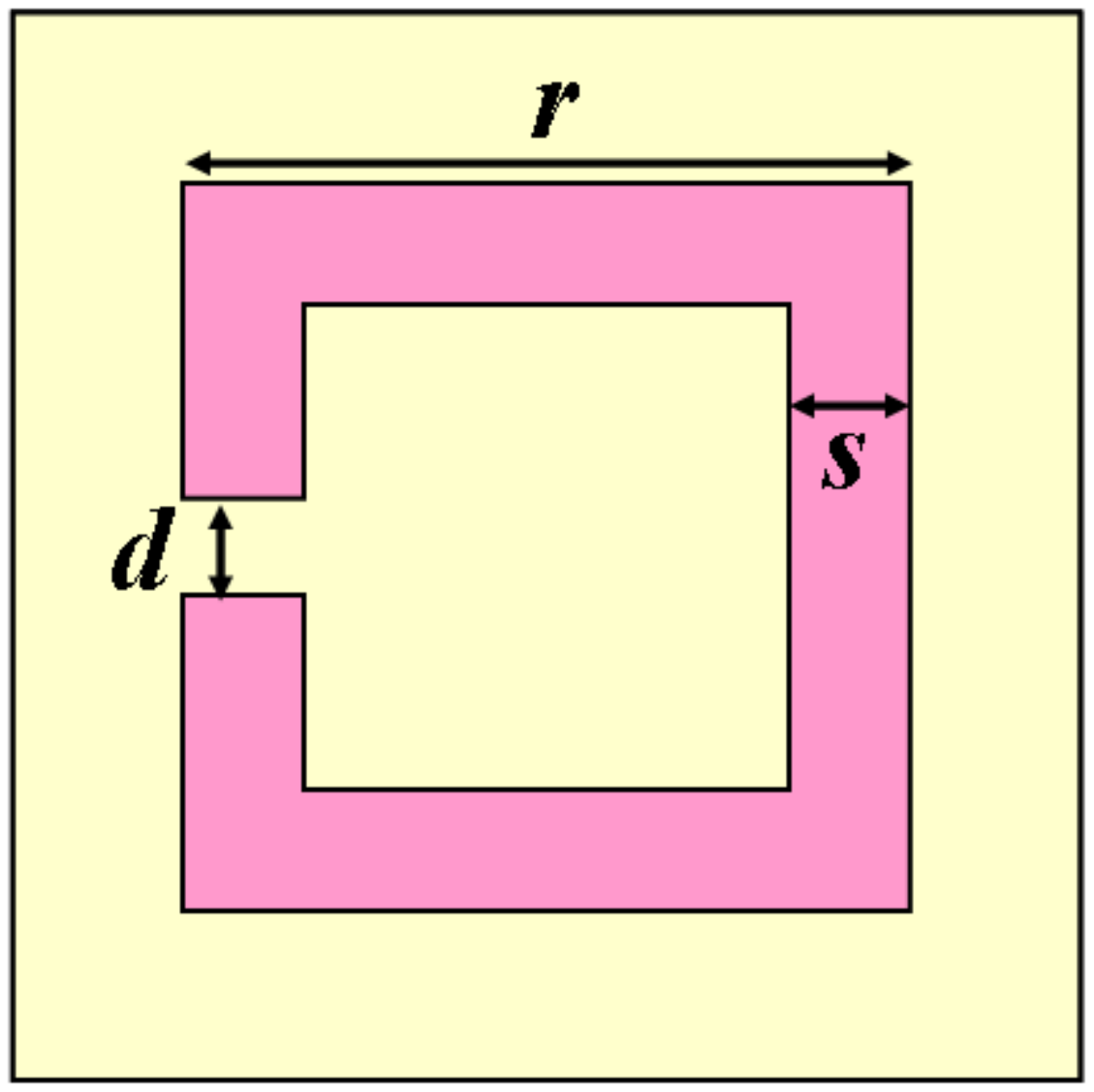}}
\caption{Unit cell configurations for a rectangular waveguide loaded with (a) ELC resonators and (b) SRR resonators. (c) ELC resonator dimensions. (d) SRR resonator dimensions. }
\label{fig2}
\end{figure}

Fig.~\ref{fig1} shows a (ELC+SRR)-loaded metallic waveguide with a coaxial excitation of TE modes.
The waveguide has $9 \times 9$ mm cross section and $70$ mm length.
The ELC resonators are placed at the lateral sides of the waveguide and the SRR resonators are placed at the center of the waveguide. Fig.~2(a, b) show unit cells of ELC-loaded and SRR-loaded waveguides, respectively. 
In the (ELC+SRR)-loaded waveguide, the ELC and SRR resonators as shown are simply combined into one unit cell. {The unit cell dimensions are $9 \times 9$ mm in the $x$ and $y$ directions and $8$ mm in the $z$ (propagation) direction.} Referring to Fig.~2(c, d), the dimensions of the ELC resonator are as follows: $l=6$ mm, $w=0.8$ mm, $g=0.4$ mm, and $t=2.4$ mm. Similarly, for the SRR resonator we have $r=4$ mm, $s=1$ mm and $d=0.5$ mm. The resonators are backed by a dielectric substrate with $\epsilon_r=2.2$ and loss tangent $\text{tan} \, \delta=9 \times 10^{-4}$. The thickness of the substrate is 0.5 mm and the Cu metallization thickness of the resonators is 0.030 mm. The inner wire of the coaxial feed has 0.5 mm radius and it extends 8 mm inside the waveguide. The coaxial cable has 1.65 mm outer radius and dielectric constant of 2.1. All simulations are performed using the commercial software CST Microwave Studio$^{\text{TM}}$ (MWS), based on the finite integration technique. The dispersion diagrams are calculated through the eigensolver of MWS, with a periodic boundary condition (except for the phase shift) assigned along the $z$ direction.
\begin{figure}[htbp]
\centering
\includegraphics[width=0.78\textwidth]{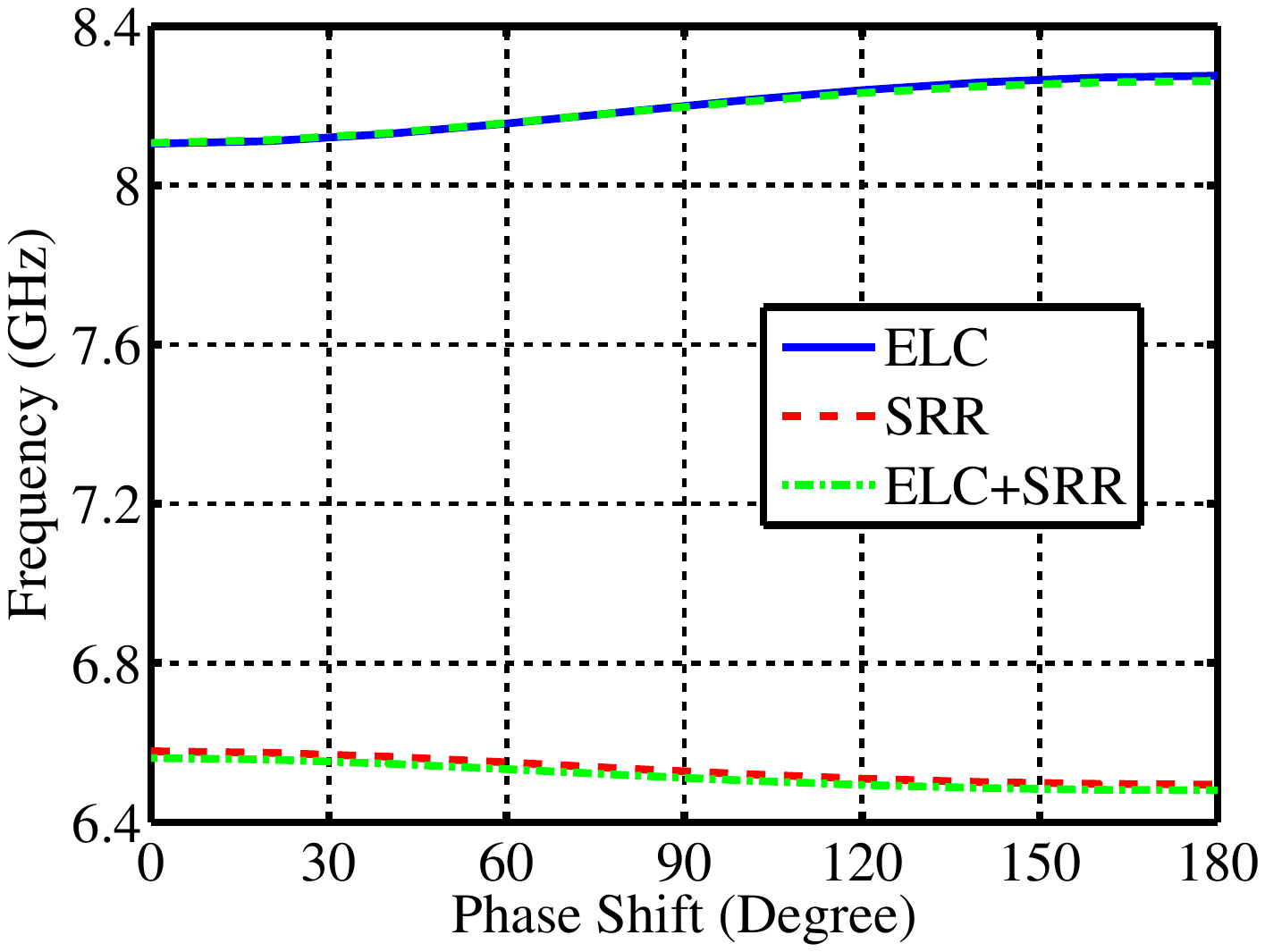}
\caption{TE-mode dispersion diagrams of ELC-, SRR-, and (ELC+SRR)-loaded waveguides. 
The SRR-loaded waveguide supports backward waves below cutoff, whereas the ELC-loaded waveguide supports forward waves. The first and second passbands of the (ELC+SRR)-loaded waveguide nearly match the passbands of the SRR-loaded and ELC-loaded waveguides, respectively.}
\label{fig3}
\end{figure}
\begin{figure}[htbp]
\centering
\includegraphics[width=0.78\textwidth]{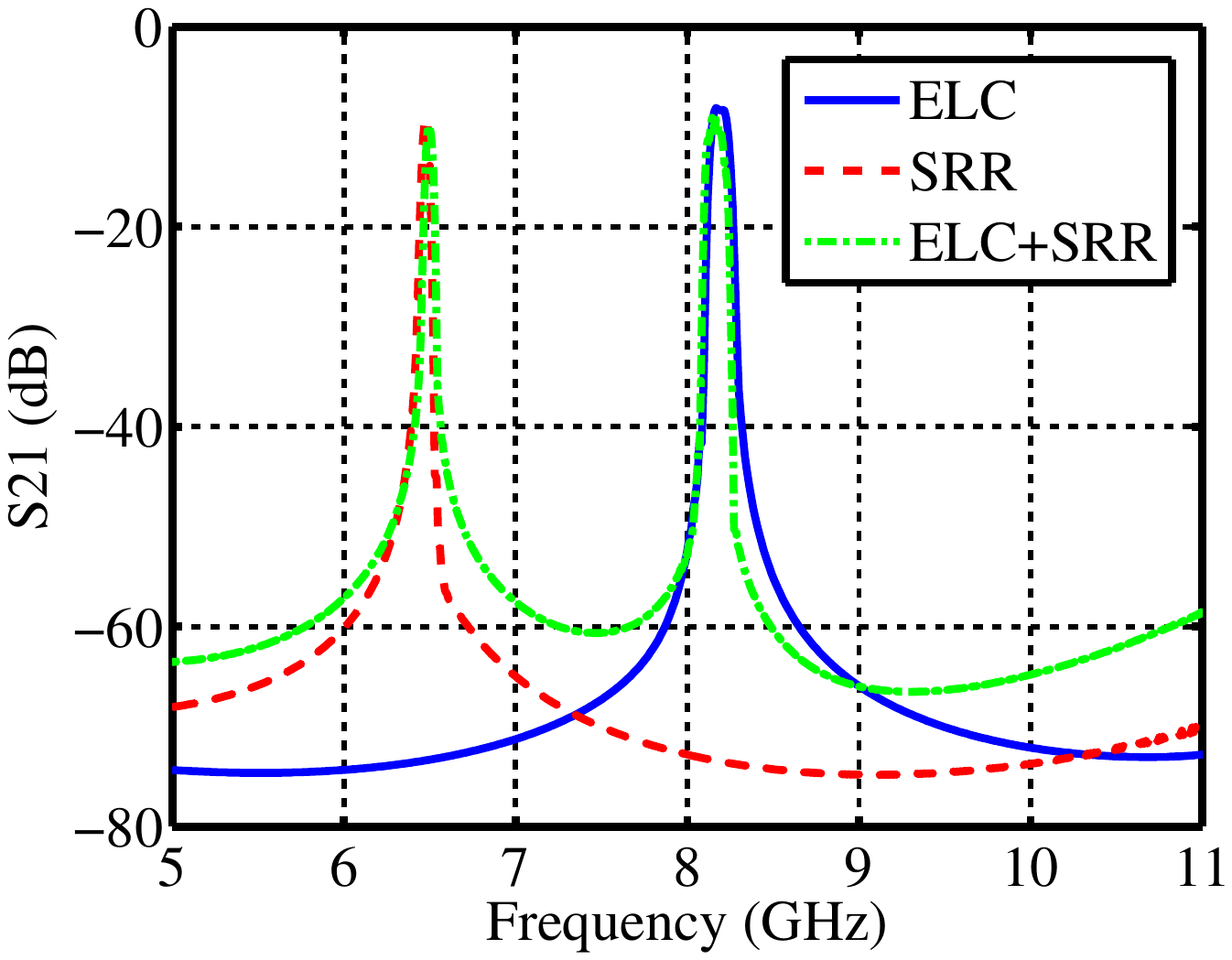}
\caption{TE-mode transmission spectra for ELC-, SRR-, and (ELC+SRR)-loaded waveguides. For the (ELC+SRR)-loaded waveguide, two transmission peaks nearly match the peaks of the SRR-loaded and ELC-loaded waveguides, respectively. The transmission peaks have similar amplitudes.}
\label{fig4}
\end{figure}
\begin{figure}[htbp]
\centering
\includegraphics[width=0.78\textwidth]{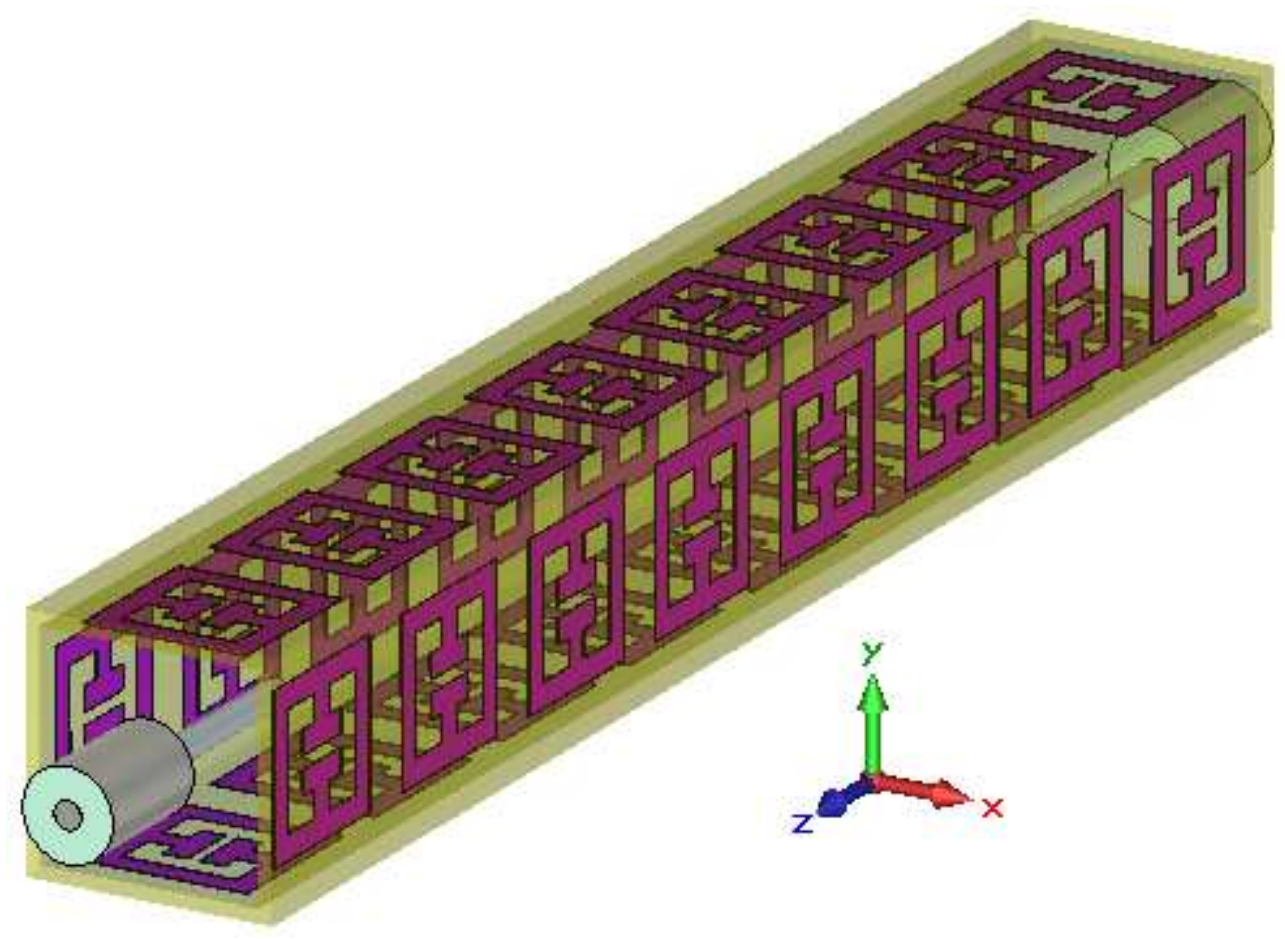}
\caption{Rectangular waveguide loaded with ELC resonators placed on all four side-walls, under TM mode excitation. Two orientations are considered for the ELC resonators. The first orientation is shown in the figure. The second orientation has the lateral-walls ELC resonators rotated by $90^o$ along the $x$ direction and the top/bottom-walls ELC resonators rotated by $90^o$ along the $y$ direction. }
\label{fig5}
\end{figure}
\begin{figure}[htbp]
\centering
\includegraphics[width=0.78\textwidth]{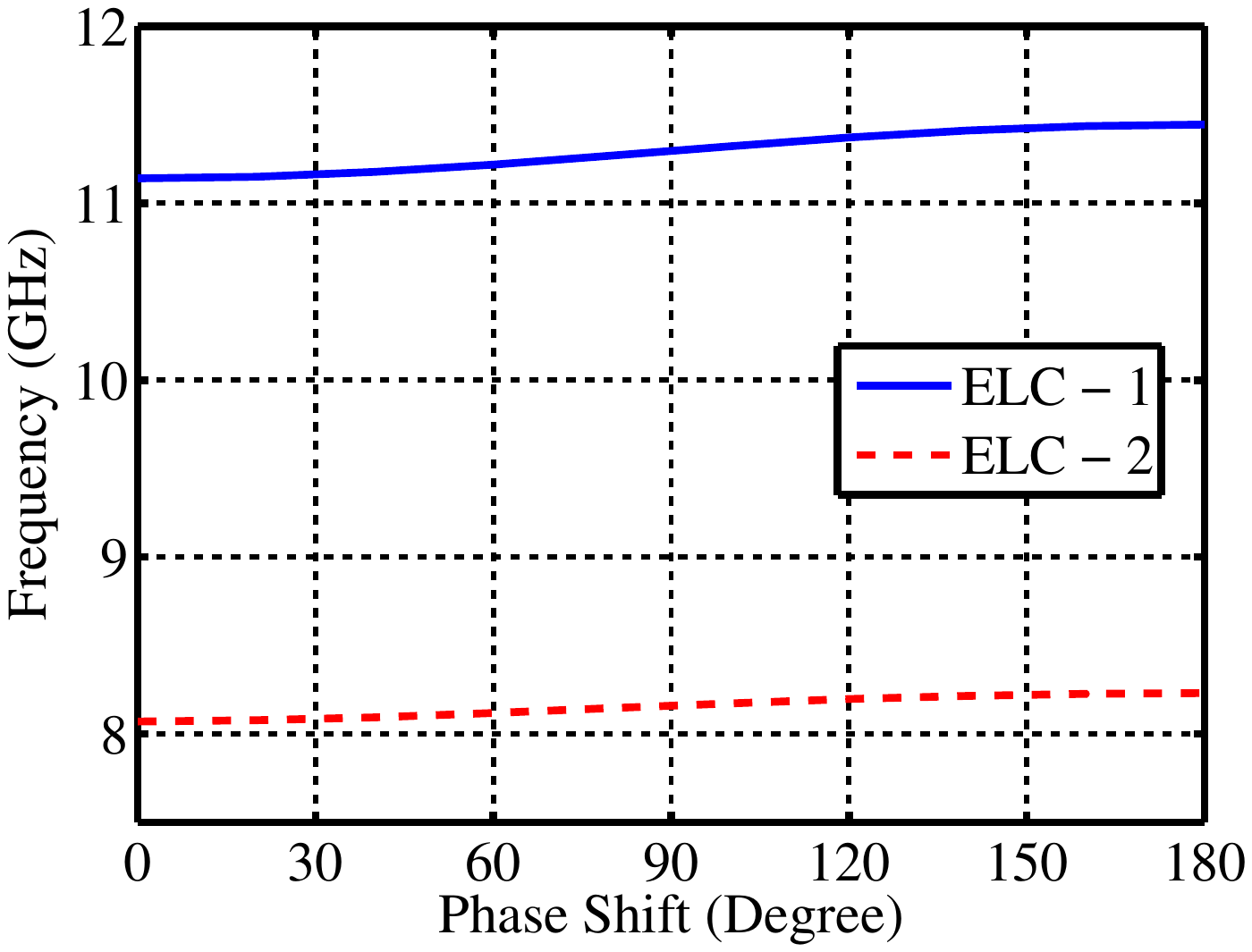}
\caption{TM-mode dispersion diagrams of ELC-loaded waveguides with the two ELC resonator orientations considered.}
\label{fig6}
\end{figure}
\begin{figure}[htbp]
\centering
\includegraphics[width=0.78\textwidth]{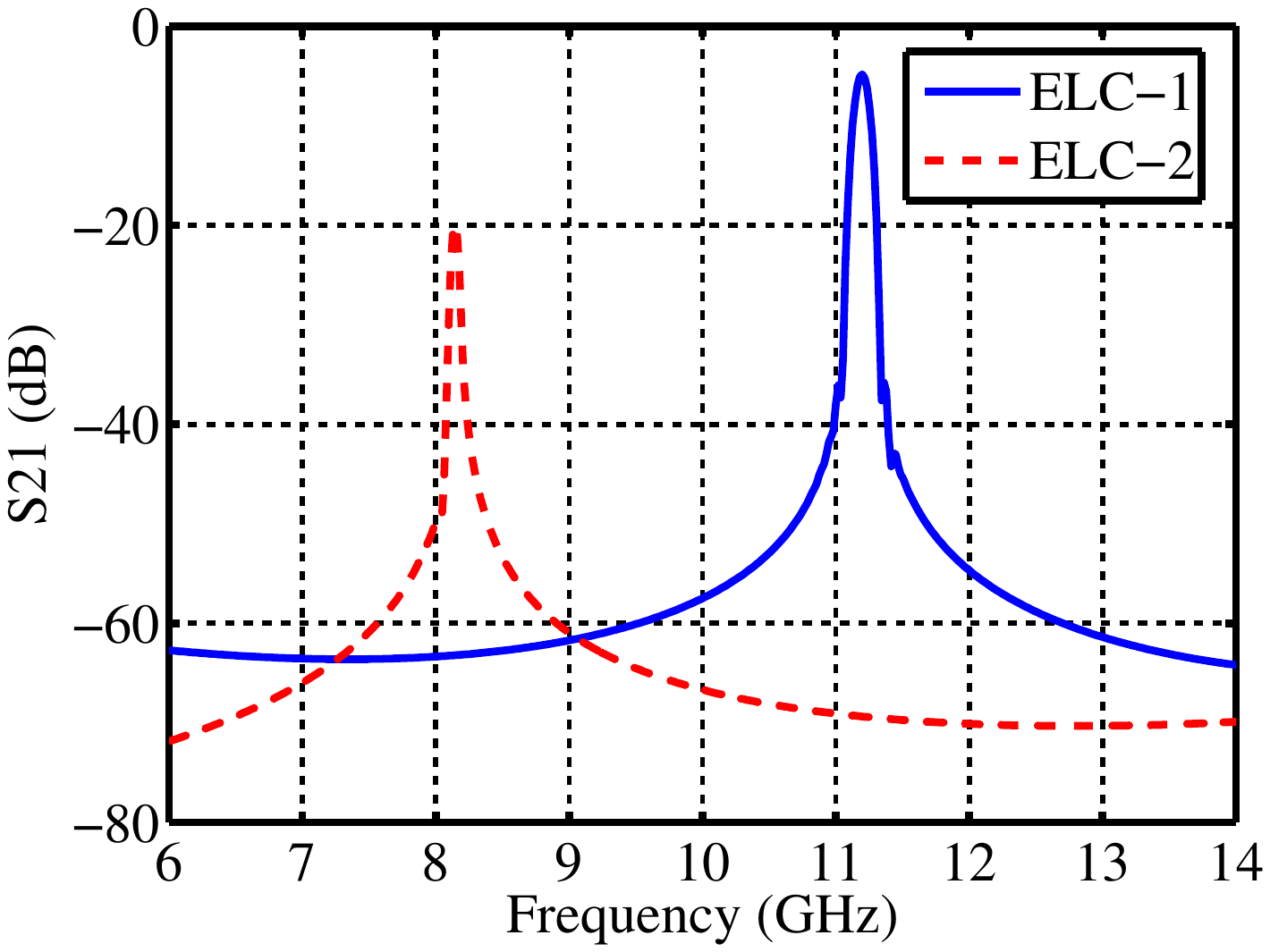}
\caption{TM-mode transmission spectra of ELC-loaded waveguides with the two ELC resonator orientations considered.}
\label{fig7}
\end{figure}

Fig.~\ref{fig3} shows the TE-mode dispersion diagrams for ELC-, SRR- and (ELC+SRR)-loaded waveguides, respectively. For the SRR-loaded waveguide, we observe a negative slope characteristic of a backward wave, whereas for the ELC-loaded waveguide we observe a positive slope characteristic of a forward wave. When the waveguide is loaded with both SRR and ELC resonators, the first passband virtually overlaps with that of the SRR-loaded waveguide and the second passband virtually overlaps with that of the ELC-loaded waveguide, showing that the passband of the backward waves and forward waves can be controlled independently. Fig.~\ref{fig4} shows the corresponding transmission spectra, which match very well the dispersion diagrams.

Fig.~\ref{fig5} shows another ELC-loaded waveguide configuration, now with coaxial excitation of TM modes. The waveguide has $9 \times 9$ mm cross section and $64$ mm length. In this case, we insert ELC resonators at the four walls in view of the TM$_{11}$ mode distribution. Moreover, we simulate two different ELC orientations: The first orientation (ELC-1) is identical as before whereas the second orientation (ELC-2) has the ELC resonators rotated by a $90^o$ angle (remaining parallel to the walls). Fig.~\ref{fig6} shows the TM-mode dispersion diagrams for these two waveguides. The passband for first orientation occurs around 11.8 GHz and the passband for the second orientation occurs around 8 GHz. Both passbands correspond to forward waves. Fig.~\ref{fig7} shows the corresponding transmission spectra for both orientations. For the second orientation, the transmission peak is noticeably weaker. 

For the ELC-loaded waveguides considered, the transmission spectra of TE and TM modes are significantly different because for the TE case the electric field in parallel to the ELC resonators, whereas for the TM case the electric field is perpendicular to the ELCs. Note also that the orientation of the SRR resonators 
is chosen so that they are properly excited by a perpendicular magnetic field (TE mode).

In summary, it was shown that ELC-loaded waveguides support forward waves below the (original frequency) cutoff for both for TE and TM type of excitation. Moreover, when loading metallic waveguides with ELC and SRR resonators simultaneously, both forward and backward propagating modes can be produced below cutoff, and each mode can be independently controlled by changing the dimensions of the respective (ELC and SRR) resonators, respectively. Based on transmission spectra results, we conclude that loading by ELC resonators provide an attractive strategy for the miniaturization of metallic waveguides.

\bibliography{references}

\end{document}